\pdfoutput=1

\documentclass[11pt]{article}

\usepackage[]{acl}

\usepackage{times}
\usepackage{latexsym}

\usepackage[T1]{fontenc}

\usepackage[utf8]{inputenc}

\usepackage{microtype}

\usepackage{inconsolata}
\usepackage{booktabs} 
\usepackage{pifont}
\newcommand{\CHECK}{\textcolor{green}{\ding{51}}} 
\newcommand{\CROSS}{\textcolor{red}{\ding{55}}} 
\usepackage{graphicx} 
\usepackage{multirow}
\usepackage{multicol}
\usepackage{arydshln}
\usepackage{color}
\usepackage{array}
\usepackage{amssymb}
\usepackage{amsmath}

\usepackage{xcolor}
\usepackage{tcolorbox}
\usepackage{makecell}
\usepackage{listings}
\usepackage{subcaption}

\title{Towards Reliable Large Audio Language Model}

\author{
Ziyang Ma\textsuperscript{1},
Xiquan Li\textsuperscript{1},
Yakun Song\textsuperscript{1},
Wenxi Chen\textsuperscript{1}, 
Chenpeng Du\textsuperscript{2},
Jian Wu\textsuperscript{2}, \\
\textbf{
Yuanzhe Chen\textsuperscript{2},
Zhuo Chen\textsuperscript{2},
Yuping Wang\textsuperscript{2},
Yuxuan Wang\textsuperscript{2},
Xie Chen\textsuperscript{1,3}\footnotemark[2]} \\
$^1$ X-LANCE Lab, School of Computer Science, \\ 
MoE Key Lab of Artificial Intelligence, Shanghai Jiao Tong University \\ 
$^2$ ByteDance, $^3$Shanghai Innovation Institute
}

\begin{document}
\maketitle

\renewcommand{\thefootnote}{\fnsymbol{footnote}}
\footnotetext[2]{Corresponding author}
\renewcommand{\thefootnote}{\arabic{footnote}}

\begin{abstract}
Recent advancements in large audio language models (LALMs) have demonstrated impressive results and promising prospects in universal understanding and reasoning across speech, music, and general sound. 
However, these models still lack the ability to recognize their knowledge boundaries and refuse to answer questions they don't know proactively. 
While there have been successful attempts to enhance the reliability of LLMs, reliable LALMs remain largely unexplored. 
In this paper, we systematically investigate various approaches towards reliable LALMs, including training-free methods such as multi-modal chain-of-thought (MCoT), and training-based methods such as supervised fine-tuning (SFT). 
Besides, we identify the limitations of previous evaluation metrics and propose a new metric, the Reliability Gain Index (RGI), to assess the effectiveness of different reliable methods. 
Our findings suggest that both training-free and training-based methods enhance the reliability of LALMs to different extents. 
Moreover, we find that awareness of reliability is a ``meta ability'', which can be transferred across different audio modalities, although significant structural and content differences exist among sound, music, and speech. 
\end{abstract}

\section{Introduction}
\begin{figure}[htbp]
    \centering
    \includegraphics[width=1\linewidth]{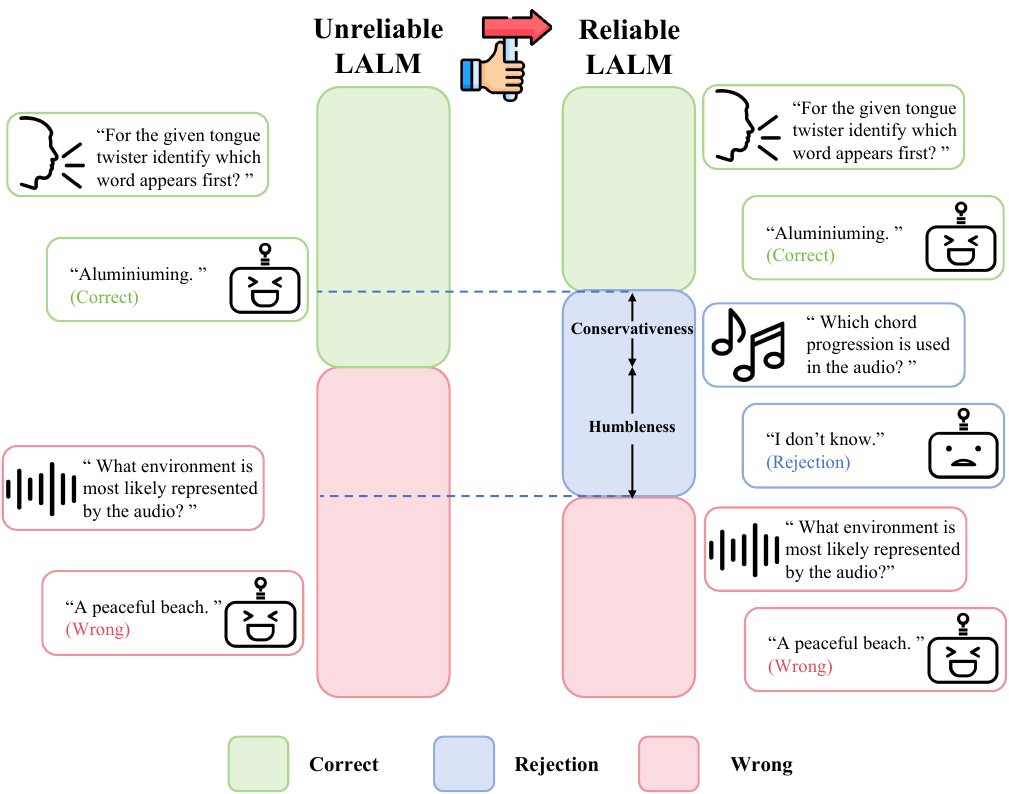}
    \caption{Illustration of Unreliable vs. Reliable LALMs. }
    \label{fig:ReliableLALM}
\end{figure}
Large audio language models (LALMs) have emerged as a promising approach to address the complex challenges of universal understanding and reasoning across diverse audio modalities, including speech~\cite{wang2023blsp, hu2024wavllm, deng2024wav2prompt}, music~\cite{deng2023musilingo, liu2024music}, and general sound~\cite{gong2023listen, kong2024audio}. 
LALMs leverage the power of large-scale pre-trained encoders and LLMs to capture intricate acoustic features and semantic representations across various scenarios~\cite{gong2023joint, ghosh2024gama, tang2023salmonn, chu2023qwen, Qwen2-Audio}, demonstrating potential to handle a wide range of unsolved tasks. 
Despite their impressive performance, LALMs still face a significant limitation: they lack the ability to recognize when they do not know the answer to a question, as shown in Figure~\ref{fig:ReliableLALM}. 
The left part depicts an unreliable LALM, where the model provides either correct or incorrect answers for audio inputs (sound, music, or speech) without rejecting awareness. The right part shows a reliable LALM, where the model refuses to answer questions that exceed its knowledge boundary. 
This reliability helps prevent models from offering incorrect or overly confident responses when faced with uncertainty. 

While there have been successful efforts to enhance the reliability of language models in text-based models~\cite{yang2023alignment, cheng2024can, yona2024can, zhang2023r, xu2024rejection}, reliable LALMs remain largely unexplored. 
Building reliability in LALMs is crucial for applications where the model’s confidence is essential, especially in real-world scenarios such as healthcare~\cite{jia2024medpodgpt}, autonomous driving~\cite{xu2024drivegpt4}, and interactive agents~\cite{ma2024language}. 
A reliable LALM should not only provide accurate answers but also have the ability to refuse to answer when it is unsure, offering a more responsible interaction model. 

In this paper, we present a comprehensive investigation into methods for enhancing the reliability of LALMs. 
We explore both training-free and training-based approaches that can effectively improve LALM's ability to identify its knowledge boundaries and reject incorrect answers. 
Specifically, we investigate how these methods influence accuracy, truthfulness, and reliability in a variety of audio modalities. 
Furthermore, we introduce a novel quantitative metric, the Reliability Gain Index (RGI), to assess the effectiveness of different reliability-enhanced techniques. 
The goal is for the model to avoid being overly conservative (from what it knows), while promoting humility (from what it doesn't konw) by rejecting answering, as shown in Figure~\ref{fig:ReliableLALM}. 
Our experiments demonstrate that the ability to say “I don’t know” (IDK) is a ``meta ability'' of LALMs, which means this ability can be trained on one modality and transferred to other audio modalities, even in the presence of significant structural and content differences among sound, music, and speech. 
Our contribution can be summarized in the following points:
\begin{enumerate}
    \item To the best of our knowledge, we are the first to investigate the reliability of LALM. Both training-free and training-based methods are conducted to verify LALM's reliability. 
    \item We identify the limitations of previous evaluation metrics and propose a new metric, the Reliability Gain Index (RGI), to measure the effectiveness of different reliable methods. 
    \item With our new metric under the cross-modal setting, we find that the awareness of reliability is a ``meta ability'' and can be transferred to other modalities in the context of large audio language modeling. 
\end{enumerate}

\section{Related Work}
\subsection{Large Audio Language Model}
Large audio language models (LALMs), as a rising part of multimodal large language models (MLLMs), aim to leverage the capabilities of LLMs to achieve advanced audio understanding and reasoning abilities. 
However, the inherent differences in acoustic structures and content across speech, music, and general sound make universal audio processing challenging, primarily due to the domain conflict~\cite{wang2023lauragpt} and the catastrophic forgetting~\cite{tang2023salmonn} problem. Balancing the performance across these different modalities remains difficult. 

Some works focus on modeling individual modalities with LLMs among speech, music, and general sound. 
In speech language models (SLMs), tasks such as LLM-based speech recognition (ASR)~\cite{li2023prompting, wu2023decoder, ma2024embarrassingly, yu2023connecting, yang2024mala, yang2024ctc, geng2024unveiling, ma2025speech}, speaker diarization (SD)~\cite{shi2024advancing, meng2024large}, and speech emotion recognition (SER)~\cite{xu2024secap, lin2024paralinguistics, cheng2024emotion, kang2024frozen} are critical, which involve either acoustic features, semantic features, or both, specific to speech. Furthermore, some works tackle various tasks with all-in-one modeling within the speech domain~\cite{wang2023blsp, hu2024wavllm, deng2024wav2prompt}. 
In the realm of LLM-based music understanding, some approaches focus on solving problems related to signal-based music processing~\cite{deng2023musilingo, liu2024music}, while others target symbolic music understanding~\cite{yuan2024chatmusician, qu2024mupt}. 
Within general sound understanding, LLM-based automated audio captioning (AAC)~\cite{wu2023beats, chen2024slam, li2024drcap, liu2024leveraging} and subsequent audio question answering (AQA) task~\cite{gong2023listen, kong2024audio} have received significant attention due to their potential to advance the field. 
Further research has explored methods that can handle two~\cite{gong2023joint, ghosh2024gama} or three~\cite{tang2023salmonn, chu2023qwen, Qwen2-Audio} modalities simultaneously by scaling data and model parameters, as well as innovating in the design of data pipelines~\cite{lu2024desta, lu2024developing} and model architectures~\cite{bhati2024state}. 

The recently developed open-source model, Qwen2-Audio~\cite{Qwen2-Audio}, demonstrates strong general-purpose audio understanding and reasoning capabilities across various benchmarks~\cite{yang2024air, li2024omnibench, sakshi2024mmau} through a combination of pre-training, supervised fine-tuning (SFT), and reinforcement learning with human feedback (RLHF). Despite these advances, even Qwen2-Audio still lacks awareness of reliability in our experiments. This highlights the need for further exploration into the reliable LALM, an area that remains an open challenge in the field. 

\subsection{Evaluation of Reliable Generation}
Evaluating the reliability of LLMs is a new topic that lacks a unified standard. In general, the goal is for LLMs to be able to refuse answering when they are unable to derive an answer. Several approaches have been proposed to assess the reliability of LLMs.
\citet{yang2023alignment} introduced the use of Prudence Score, Over-Conservativeness Score, and Honesty Score to evaluate model performance. 
\citet{cheng2024can} were the first to introduce the concept of the IDK dataset and employed Knowledge Quadrants to visualize a model’s knowledge coverage and then defined the Truthfulness score.
\citet{yona2024can} defined the Mean Faithful Generation (MFG) metric, which quantifies the expected faithfulness of a single model output by comparing it against a ground truth. 
Additionally, \citet{zhang2023r} proposed the Average Precision (AP) score measuring the model's precision in identifying and ranking relevant predictions based on its knowledge. 
A more recent and popular method for evaluating reliability involves weighting accuracy and truthfulness to obtain a final reliability score introduced by ~\citet{xu2024rejection}. 
This method provides a more holistic measure of a model’s overall reliability by balancing both the model's correct responses and its ability to reject uncertain answers. 

\section{Reliable LALM}
The core purpose of reliable LALM is to refuse to answer when the model doesn't know the answer given the input audio and instruction. 
We explored two distinct approaches to enhance the model's reliability: training-free and training-based methods. 
The training-free method aims to activate the inherent capabilities of the model through prompting or agent, thereby achieving reliability without requiring additional training. 
In contrast, the training-based method seeks to enhance the model's reliability by post-training, thereby explicitly advertising reliability through the training process. 

\subsection{Training-free Method}
\label{sec:training_free_method}
Three training-free methods for enhancing model reliability are employed: IDK Prompting, MCoT Prompting, and Task Agent. These methods leverage the model's inherent instruction-following capabilities to improve its reliability without additional training. 

\paragraph{IDK Prompting.}
I Don’t Know (IDK) Prompting is a method designed to enhance model reliability by adding a supplementary prompt after the input question. This prompt encourages the model to acknowledge uncertainty in cases where it lacks sufficient information to provide a confident answer. By incorporating this strategy, the model is prompted to explicitly state ``I don’t know'' when necessary. The specific prompt used in this method is outlined in Appendix~\ref{sec:prompt_template_for_LALM}. 

\paragraph{MCoT Prompting.}
Multi-modal Chain-of-Thought (MCoT) Prompting~\cite{lu2022learn, zhang2023multimodal} encourages the LALM to reason step by step, with the intention of refining its analysis of the given problem and producing more reliable results. This approach leads the model to break down complex tasks into smaller, manageable components, which are processed sequentially. By prompting the model to articulate its thought process, MCoT Prompting improves its ability to reason logically and arrive at reliable conclusions. The specific prompt for MCoT is provided in Appendix~\ref{sec:prompt_template_for_LALM}.

\paragraph{Task Agent.}
Audio data differs from textual data significantly, presenting substantial differences in the content and structure among speech, sound, and music, despite all being categorized as ``audio''. 
We propose to use a task agent, which is designed to support multi-step reasoning with tool-using ability during the inference process. 
This approach incorporates a sequence of steps for reasoning that includes identifying the type of audio, analyzing its content, and producing a final prediction. The steps are as follows:
\begin{enumerate}
    \item \textbf{Identify the type of audio.} The model is first required to categorize the audio input. The possible types include sound, music, and speech. 
    \item \textbf{Generate content based on audio type.} Depending on the identified audio type, the model generates the corresponding content. For speech, the model outputs the automatic speech recognition (ASR) result. For sound or music, the model generates an automatic audio caption (AAC) or music caption (MC). 
    \item \textbf{Output generation.} After completing the previous steps, the model combines the audio, question, and generated content, and then inputs them into the LALM to obtain the final answer. 
\end{enumerate}
This multi-step reasoning process ensures that the model can make context-aware decisions, further enhancing its reliability. We provide the detailed prompt processing in Appendix~\ref{sec:prompt_template_for_LALM}.

\subsection{Training-based Method}
Following previous works~\cite{yang2023alignment, cheng2024can, zhang2023r}, we adopt a similar approach but apply it to the multi-modal setting. The training-based method involves two key steps: construction of a model-specific IDK dataset and post-training of the model. 

\paragraph{Construction of the IDK Dataset.}
Given that different models possess varying knowledge quadrants, it is essential to construct a model-specific IDK dataset for each model. 
For each data point, $N$ possible answers are sampled. If the model provides the correct answer at least $K$ times (where $K \leq N$), we assume the model's knowledge adequately covers the question, and the original answer is retained as the ground truth. Conversely, if the model fails to answer correctly enough times, the answer is labeled as IDK. 
The threshold parameter, denoted as $K$@$N$, plays a critical role in defining the IDK dataset.
$0$@$N$ means no IDK data is generated, where all answers retain their original labels. 
While $N$@$N$ means the model must answer all $N$ times correctly to retain the original label. 
By adjusting this threshold, different knowledge levels of the model-specific IDK dataset can be generated. 

\paragraph{Post-training with the IDK Dataset.}
Once the IDK dataset is constructed, we perform supervised fine-tuning (SFT) to align the model’s reliability. Specifically, we utilize the IDK dataset to fine-tune the model, guiding it to better handle uncertainty and enhance its reliability. During this process, the model learns to recognize when it should confidently provide an answer and when it should appropriately output IDK. 

\begin{figure*}[htbp]
    \centering
    \includegraphics[width=1\linewidth]{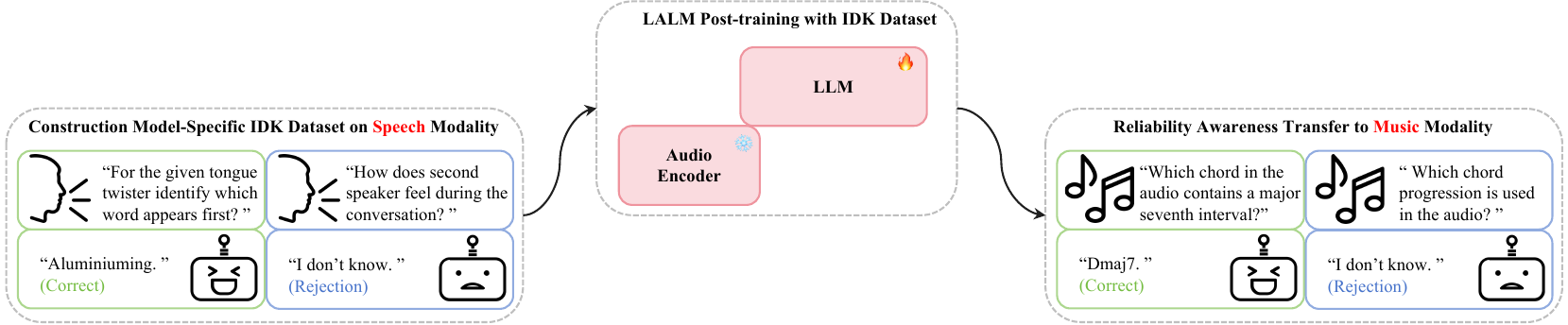}
    \caption{Illustration of Reliability Awareness Transferability in LALM. }
    \label{fig:ReliabilityTransfer}
\end{figure*}

\section{Evaluation}
We first introduce a basic evaluation metric for truthfulness proposed by \citet{cheng2024can} and further, reliability proposed by \citet{xu2024rejection}.  We then propose our new metric, Reliability Gain Index (RGI), to measure the effectiveness of different reliable methods. 

\subsection{Reliability Evaluation}
To evaluate the reliability of the given LALM, we consider using the weighted overall reliability that balances the model’s helpfulness and truthfulness. 
Let $N$ denote the total number of queries tested, which can be expressed as: 
\begin{equation}
N = N_c + N_r + N_w,
\end{equation}
where $N_c, N_r, N_w$ donate the numbers of correct, rejected, and wrong answers. 
Accuracy (Acc) measures the proportion of correct answers relative to the total number of queries. It is calculated as: 
\begin{equation}
Acc = \frac{N_c}{N}. 
\end{equation}
Truthfulness (Tru) quantifies how truthful the model's answers are when it does not reject a query. It is defined as the proportion of not wrong among queries, which is given by:
\begin{equation}
Tru = 1- \frac{N_w}{N}. 
\end{equation}
Rejection Rate (Rej) measures the fraction of queries for which the model chooses to refuse to answer. It is calculated as:
\begin{equation}
Rej = \frac{N_r}{N}
\end{equation}
The Rejection Rate reflects the model’s willingness to reject uncertain or out-of-scope queries, thereby avoiding providing potentially unreliable answers. 
Reliability (Rel) combines the model’s accuracy and truthfulness into a single measure. It accounts for both the model’s performance on correct answers and its ability to reject uncertain responses. The Reliability score is given by: 
\begin{equation}
\label{eq:rel}
Rel = Rej \cdot Acc + (1-Rej) \cdot Tru,
\end{equation}
where the Reliability is weighted by the rejection rate. The Rejection Rate represents the degrees of sensitivity towards errors. 

\subsection{Reliability Gain}
While the previously defined metrics such as Accuracy, Truthfulness, Rejection Rate, and Reliability are useful for evaluating the absolute capability of a model to express IDK, they are less effective in measuring the relative effectiveness of different reliable methods. 
Specifically, these metrics fail to reveal how well a method balances two crucial aspects of reliability: conservativeness (the tendency to reject correct answers) and humbleness (the tendency to reject incorrect answers).

To capture the effectiveness of reliable methods more precisely, we introduce the Reliability Gain Index (RGI). 
Specifically, the RGI evaluates the relative gains in the model's rejection capabilities, distinguishing between increases in relative conservativeness and relative humbleness. 
Let $N_c$ and $N_w$ denote the numbers of correct and incorrect answers, respectively, in the original unreliable model: 
\begin{equation}
N = N_c + N_w. 
\end{equation}
After applying a reliable method (e.g., prompting or supervised fine-tuning), the numbers of correct and wrong answers may be redistributed into different categories:
\begin{equation}
\begin{aligned}
N_c &= N_{cc} + N_{cr} + N_{cw}, \\
N_w &= N_{wc} + N_{wr} + N_{ww},
\end{aligned}
\end{equation}
where $N_{cc}$ refers to correct answers that remain correct, $N_{cr}$ refers to correct answers that are rejected, and $N_{cw}$ refers to correct answers that are reclassified as wrong. The same marks are also applied to $N_w$. 

To quantify how much the model has become more conservative, we define the relative conservativeness increase as:
\begin{equation}
\label{eq:delta_con}
\Delta_{Con} = \frac{N_c - N_{cc}}{N_c}. 
\end{equation}
This metric captures the proportion of correct answers that were rejected after applying the reliable method. A higher value of $\Delta_{Con}$ indicates that the model has become more conservative in its response, rejecting more previously correct answers.
To measure the increase in the model's humbleness, we define the relative humbleness increase as: 
\begin{equation}
\label{eq:delta_hum}
\Delta_{Hum} = \frac{N_w - N_{ww}}{N_w}
\end{equation}
This metric reflects the proportion of wrong answers that were correctly rejected (i.e., converted into IDK). A higher value of $\Delta_{Hum}$ indicates that the model has become more humble by rejecting answers beyond its knowledge or capacity. 
Finally, we introduce the Reliability Gain Index (RGI) to combine the changes in conservativeness and humbleness. The RGI is defined as: 
\begin{equation}
RGI = \log(\frac{\Delta_{Hum}}{\Delta_{Con}})
\end{equation}
This metric provides a measure of the model’s improvement in reliability. A higher RGI value reflects that the model has become more humble while avoiding excessive conservatism, as it demonstrates a favorable index between increasing the rejection of wrong answers and minimizing the rejection of correct ones. 
For a concrete example illustrating why traditional metrics may inadequately capture the effectiveness of reliability methods, and a formal derivation of the conditions under which these metrics may be invalid, refer to Appendix~\ref{sec:condition_reliability}. 

\begin{table*}[h]
\centering
\caption{\textit{Accuracy} (Acc\%$\uparrow$), \textit{Truthfulness} (Tru\%$\uparrow$), and \textit{Reliability} (Rel\%$\uparrow$) performance comparison of Qwen2-Audio-7B-Instruct baselines, training-free methods, and training-based methods on the MMAU benchmark across sound, speech, and music modalities. The result for LoRA Fine-tuning is computed by cross-validation across three modalities. The best-performing items are highlighted in \textbf{bold}, and the second-best items are \underline{underlined}. We also show random guess, most frequent choice, and human evaluation results from the original MMAU paper for reference. }
\label{tab:main_results}
\resizebox{\linewidth}{!}{
\begin{tabular}{lccccccccccccc}
\toprule \toprule
\multirow{2.5}{*}{\textbf{Methods}} & \multirow{2.5}{*}{\makecell{\textbf{Post} \\ \textbf{Training}}} & \multicolumn{3}{c}{\textbf{Sound}} &  \multicolumn{3}{c}{\textbf{Music}} &  \multicolumn{3}{c}{\textbf{Speech}} & \multicolumn{3}{c}{\textbf{Total}} \\ 
\cmidrule{3-14}
& & Acc\% & Tru\% & Rel\% & Acc\% & Tru\% & Rel\% & Acc\% & Tru\% & Rel\% & Acc\% & Tru\% & Rel\% \\
\midrule \midrule
\multicolumn{14}{l}{\textbf{\textit{Baseline}}} \\ 
\midrule \midrule
MMAU (Unnormalized) & - & 54.95 & 54.95 & 54.95 & 50.98 & 50.98 & 50.98 & 42.04 & 42.04 & 42.04 & 49.20 & 49.20 & 49.20 \\
Ours (Normalized) & - & \underline{60.96} & 60.96 & 60.96 & \textbf{55.09} & 55.09 & 55.09 & \textbf{50.75} & 50.75 & 50.75 & \textbf{55.60} & 55.60 & 55.60 \\
\midrule \midrule
\multicolumn{14}{l}{\textbf{\textit{Reliable LALM}}} \\ 
\midrule \midrule
IDK Prompting & \CROSS & 58.26 & \textbf{76.28} & \textbf{73.03} & \underline{54.19} & 66.77 & 65.19 & 43.84 & 58.26 & 56.18 & 52.10 & 67.10 & 64.85 \\
MCoT Prompting & \CROSS & 57.96 & 68.17 & 67.13 & 51.50 & \textbf{71.56} & \underline{67.53} & 44.74 & \underline{60.06} & 57.71 & 51.40 & 66.60 & 64.29 \\
Task Agent & \CROSS & 58.56 & \underline{72.67} & 70.68 & 53.29 & \textbf{71.56} & \textbf{68.22} & 46.25 & 59.76 & \underline{57.93} & 52.70 & \underline{68.00} & \underline{65.66} \\
LoRA Fine-tuning & \CHECK &  \textbf{61.71} & 71.77 & \underline{70.71} & 51.35 & \underline{70.66} & 66.43 & \underline{47.90} & \textbf{61.86} & \textbf{59.91} & \underline{53.65} & \textbf{68.10} & \textbf{65.68} \\
\midrule \midrule
\multicolumn{14}{l}{\textcolor{gray}{\textbf{\textit{Reference}}}} \\ 
\midrule \midrule
\textcolor{gray}{Random Guess} & \textcolor{gray}{-} & \textcolor{gray}{26.72} & \textcolor{gray}{26.72} & \textcolor{gray}{26.72}  & \textcolor{gray}{24.55} & \textcolor{gray}{24.55} & \textcolor{gray}{24.55} & \textcolor{gray}{26.72} & \textcolor{gray}{26.72} & \textcolor{gray}{26.72} & \textcolor{gray}{26.00} & \textcolor{gray}{26.00} & \textcolor{gray}{26.00} \\
\textcolor{gray}{Most Frequent Choice} & \textcolor{gray}{-} & \textcolor{gray}{27.02} & \textcolor{gray}{27.02} & \textcolor{gray}{27.02} & \textcolor{gray}{20.35} & \textcolor{gray}{20.35} & \textcolor{gray}{20.35} & \textcolor{gray}{29.12} & \textcolor{gray}{29.12} & \textcolor{gray}{29.12} & \textcolor{gray}{25.50} & \textcolor{gray}{25.50} & \textcolor{gray}{25.50} \\
\textcolor{gray}{Human} & \textcolor{gray}{-} & \textcolor{gray}{86.31} & \textcolor{gray}{86.31} & \textcolor{gray}{86.31} & \textcolor{gray}{78.22} & \textcolor{gray}{78.22} & \textcolor{gray}{78.22} & \textcolor{gray}{82.17} & \textcolor{gray}{82.17} & \textcolor{gray}{82.17} & \textcolor{gray}{82.23} & \textcolor{gray}{82.23} & \textcolor{gray}{82.23} \\
\bottomrule \bottomrule
\end{tabular}
}
\end{table*}

\subsection{Cross-modal Reliability Awareness}
An intriguing question is whether the awareness of reliability can transfer across different audio modalities, as shown in Figure~\ref{fig:ReliabilityTransfer}. 
Using the proposed RGI metric, we can easily answer this question. 
Specifically, if we train the model on one modality and test it on another, an $RGI > 0$ indicates that the model has successfully learned to reject more questions it does not know, even when tested on a different audio modality. 
Such transferability is crucial for building robust models that can operate reliably across various domains and even various modalities. 

\section{Experiments}
\subsection{Setup}
We use Qwen2-Audio-7B-Instruct~\cite{Qwen2-Audio} as the baseline model in our experiments. This model has demonstrated strong performance across various benchmarks, including AIR-Bench~\cite{yang2024air}, OmniBench~\cite{li2024omnibench}, and MMAU~\cite{sakshi2024mmau}, positioning it as one of the most powerful open-source LALMs available. We also test the performance of other LALMs, whose detailed introduction can be found in Appendix~\ref{sec:introduction_for_different_LALMs}, and respective performance are shown in Appendix~\ref{sec:performance_of_different_LALMs}.

Our experiments are conducted on the MMAU dataset, which consists of human-annotated natural language questions and answers covering three domains: speech, environmental sounds, and music. Details of the dataset can be found in Appendix~\ref{sec:dataset_details}.

For the construction of the IDK dataset, we employ a $5$@$5$ threshold, following \citet{cheng2024can}'s setting. Specifically, for each given question, we perform 5 rounds of inference with the LALM model. If the model answers correctly in all 5 rounds, we consider the model to have sufficient knowledge of the question, and the original answer is retained as the ground truth. However, if the model answers incorrectly in any of the 5 rounds, the answer is labeled as IDK. 

For the training-free methods, we employ both single-step and multi-step reasoning introduced in Section~\ref{sec:training_free_method}. The specific prompts used in these approaches are provided in Appendix~\ref{sec:prompt_template_for_LALM}. For the training-based methods, since the Qwen-Audio series does not provide fine-tuning code, we implement our fine-tuning process based on DeepSpeed~\footnote{\url{https://github.com/microsoft/DeepSpeed}}, using Low-Rank Adaptation (LoRA)~\cite{hu2021lora} with Parameter Efficient Fine-Tuning (PEFT) library~\footnote{\url{https://github.com/huggingface/peft}}. For all three modalities, we perform SFT on the model-specific IDK dataset for $1$ epoch. Detailed training hyper-parameters for each modality are provided in Appendix~\ref{sec:training_details}.

\begin{table*}[h]
\centering
\caption{\textit{Relative Conservativeness Increase} ($\Delta_{Con}\%\downarrow$), \textit{Relative Humbleness Increase} ($\Delta_{Hum}\%\uparrow$) and \textit{Reliability Gain Index} (RGI$\uparrow$) performance of Qwen2-Audio-7B-Instruct with different reliable methods on the MMAU benchmark across sound, speech, and music modalities. The result for LoRA Fine-tuning is computed by cross-validation across three modalities. The best-performing items are highlighted in \textbf{bold}, and the second-best items are \underline{underlined}. }
\label{tab:rgi_results}
\resizebox{\linewidth}{!}{
\begin{tabular}{lccccccccccccc}
\toprule \toprule
\multirow{2.5}{*}{\textbf{Methods}} & \multirow{2.5}{*}{\makecell{\textbf{Post} \\ \textbf{Training}}} & \multicolumn{3}{c}{\textbf{Sound}} &  \multicolumn{3}{c}{\textbf{Music}} &  \multicolumn{3}{c}{\textbf{Speech}} & \multicolumn{3}{c}{\textbf{Total}} \\ 
\cmidrule{3-14}
& & $\Delta_{Con}\%$ & $\Delta_{Hum}\%$ & RGI & $\Delta_{Con}\%$ & $\Delta_{Hum}\%$ & RGI & $\Delta_{Con}\%$ & $\Delta_{Hum}\%$ & RGI & $\Delta_{Con}\%$ & $\Delta_{Hum}\%$ & RGI \\
\midrule \midrule
IDK Prompting & \CROSS & 10.81 & \textbf{20.12} & \underline{0.27} & 12.87 & 20.36 & 0.20 & 15.61 & \underline{16.52} & 0.02 & 13.10 & \textbf{19.00} & 0.16 \\
MCoT Prompting & \CROSS & 11.71 & 14.41 & 0.09 & \underline{11.68} & \underline{20.96} & \underline{0.25} & 13.21 & 16.22 & 0.09 & 12.20 & 17.20 & 0.15 \\
Task Agent & \CROSS & \underline{9.61} & \underline{16.52} & 0.24 & \textbf{11.38} & \underline{20.96} & \textbf{0.27} & \textbf{9.61} & 14.11 & \underline{0.17} & \textbf{10.20} & 17.20 & \underline{0.23} \\
LoRA Fine-tuning & \CHECK & \textbf{6.91} & 15.62 & \textbf{0.36} & 12.73 & \textbf{21.11} & 0.23 & \underline{11.56} & \textbf{18.17} & \textbf{0.19} & \underline{10.40} & \underline{18.30} & \textbf{0.26} \\
\bottomrule \bottomrule
\end{tabular}
}
\end{table*}

\subsection{Reliable Methods Analysis}
Table~\ref{tab:main_results} presents the Accuracy, Truthfulness, and Reliability metrics for different reliable methods. For the baseline on the MMAU dataset, the answers are extracted with rule-based methods, which results in inaccurate judgments of the model’s output. To address this, we further utilized the \textit{GPT-4o-mini} API to regularize the answers. The prompt template for answer normalization is shown in Appendix~\ref{sec:prompt_template_for_answer_normalization}. 
From the table, it is evident that for the different reliable methods, the model's Accuracy generally decreases compared to the baseline. However, Truthfulness improves consistently, and the Reliability metric, a weighted balance between Accuracy and Truthfulness, also shows an increase, indicating an overall improvement in the model's reliability.
Notably, the training-free methods achieve high Truthfulness, but their impact on Accuracy is relatively large. This suggests that the training-free methods tend to make the model more conservative or humble, resulting in a higher rejection rate, negatively affecting the helpfulness. In contrast, the training-based methods demonstrate a better trade-off between Accuracy and Truthfulness, as they manage to improve Truthfulness while minimizing the negative impact on Accuracy. Consequently, the Reliability of the model is improved in a more balanced manner. 

Table~\ref{tab:rgi_results} presents the Relative Conservativeness Increase, Relative Humbleness Increase, and Reliability Gain Index for different reliable methods. From Table~\ref{tab:main_results}, increasing both conservativeness and humbleness will result in an improvement in model reliability. However, a greater increase in humbleness than in conservativeness is key to achieving an effective reliable method.
From Table~\ref{tab:rgi_results} we observe that, regardless of training-free or training-based methods, the RGI is greater than $0$ across all modalities, indicating that different reliable methods are generally effective. 
When analyzing the results by modality, we find that the RGI is higher for the sound and music, while it is relatively lower for the speech. This suggests that the model is more confident about what it knows and does not know in the sound and music. The use of Task Agent (i.e., ASR results) or SFT helps mitigate this issue in speech, improving the model’s reliability. 
From all these results, it is evident that training-based methods strike a better trade-off between conservativeness and humbleness, thereby achieving a superior RGI compared to the training-free methods. 

\begin{figure}[htbp]
  \centering
  \includegraphics[width=1\linewidth]{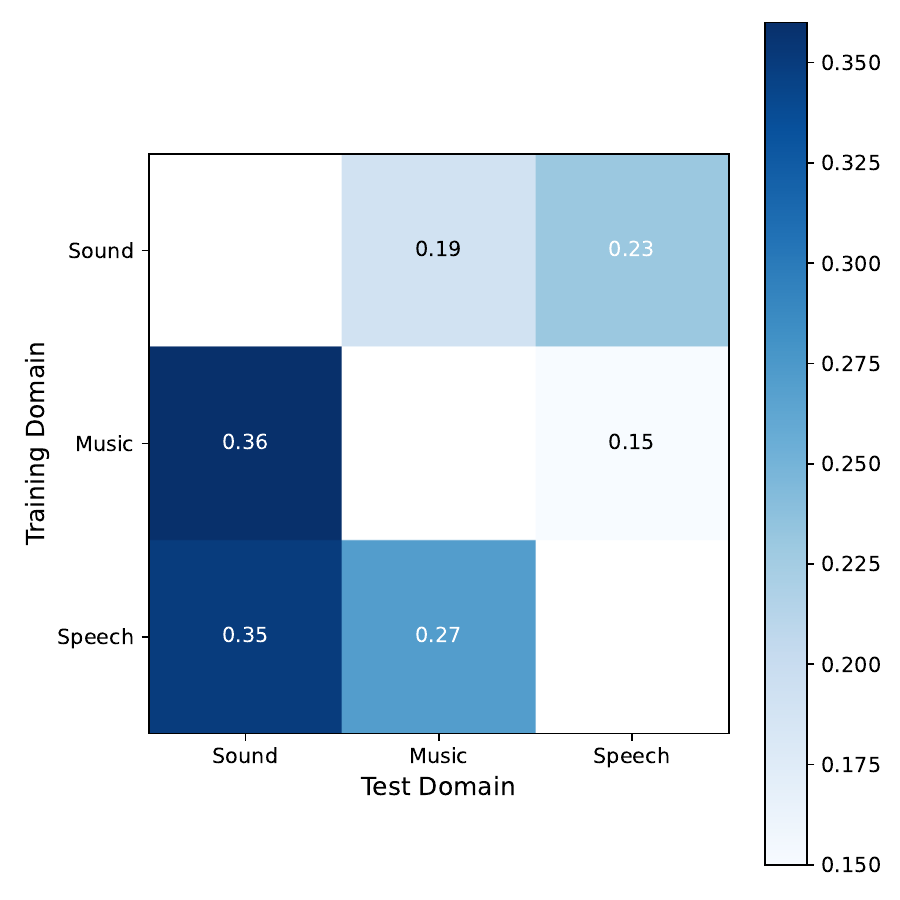}
  \caption{Heatmap for cross-modal SFT results. The figure shows the RGI
   performance of reliable models trained on one modality and tested on another. }
  \label{fig:rgi}
\end{figure}

\begin{figure*}[htbp]
    \centering
    \begin{subfigure}[b]{0.33\textwidth}
        \centering
        \includegraphics[width=\textwidth]{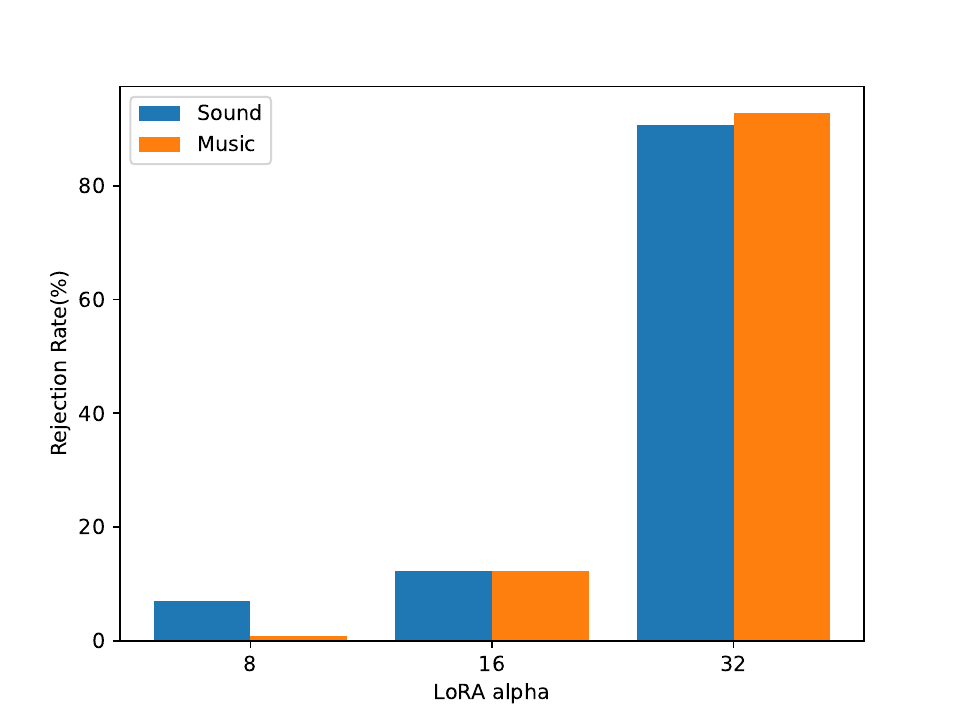}
        \caption{}
    \end{subfigure}%
    \hfill
    \begin{subfigure}[b]{0.33\textwidth}
        \centering
        \includegraphics[width=\textwidth]{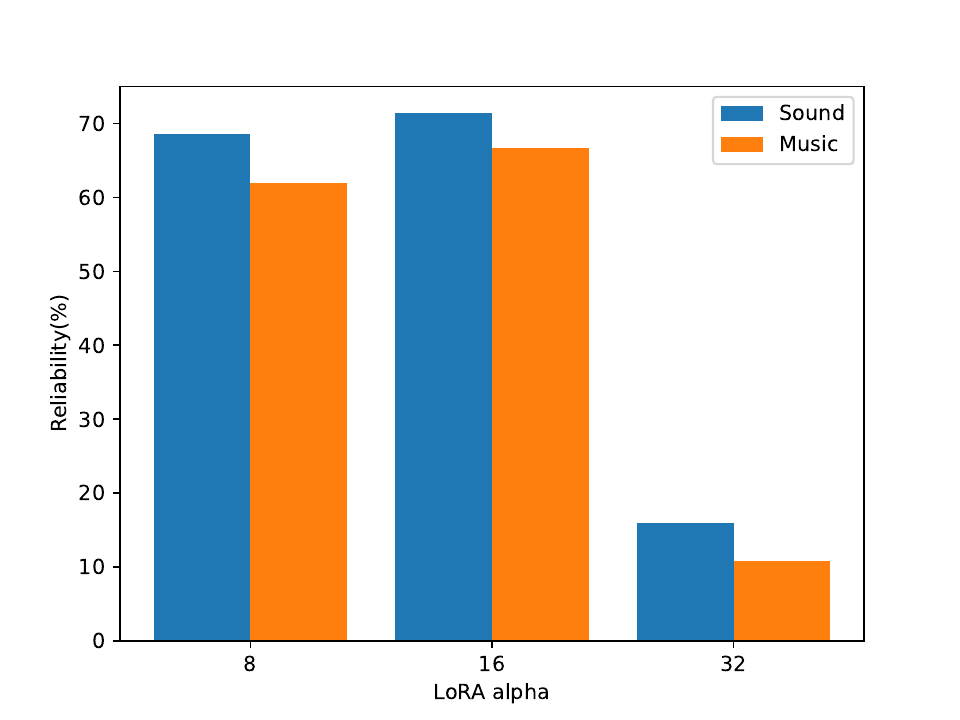}
        \caption{}
    \end{subfigure}%
    \hfill
    \begin{subfigure}[b]{0.33\textwidth}
        \centering
        \includegraphics[width=\textwidth]{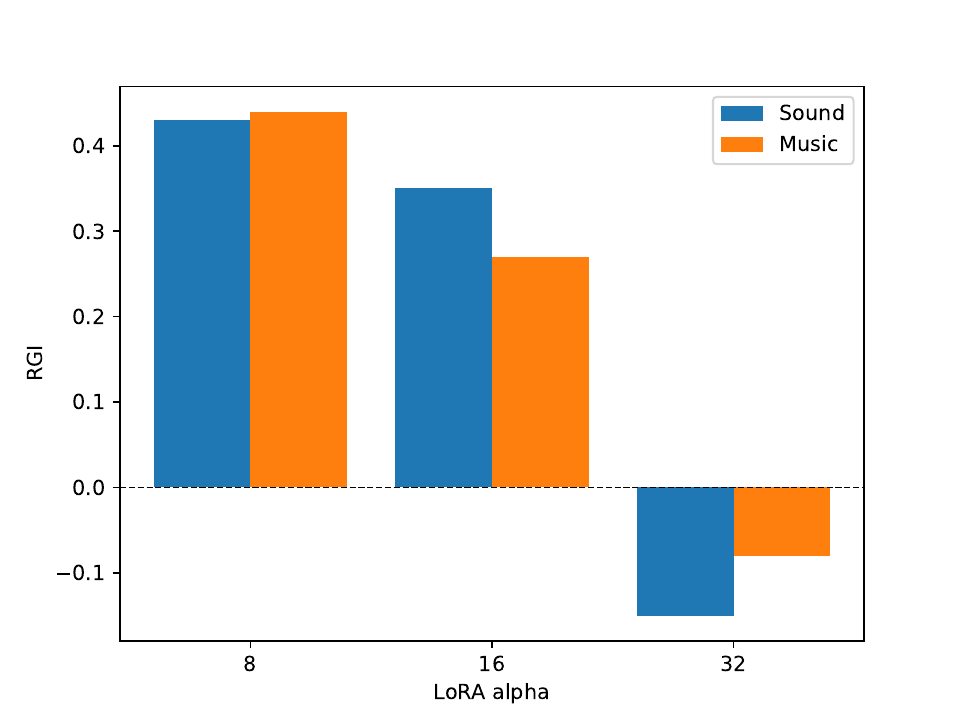}
        \caption{}
    \end{subfigure}
    \caption{Rejection Rate(\%), Reliability(\%), and Reliability Gain Index (RGI) performance for different LoRA alpha weights trained on speech modality. }
    \label{fig:lora_alpha}
\end{figure*}

\subsection{Cross-modal Analysis}
Figure~\ref{fig:rgi} presents the results of cross-model SFT. Despite the significant structural and content differences among the sound, music, and speech modalities in audio processing, the heatmap reveals that all RGI values are greater than 0. This indicates that the LALM's ability to express ``I don't know'' is a ``meta ability'', which can be learned in one modality and transferred to others.
Notably, training on one modality and testing on another often results in a high RGI when tested on the sound modality, suggesting that the model’s knowledge is more distinctly separable on sound compared to the other modalities. 
This implies that sound tasks provide clearer boundaries for what the model knows and does not know. 
Detailed cross-modal testing results can be found in Appendix~\ref{sec:LoRA_fine-tuning_results_on_different_domains}. 

\subsection{Ability Study}
\begin{figure}[htbp]
  \centering
  \includegraphics[width=1\linewidth]{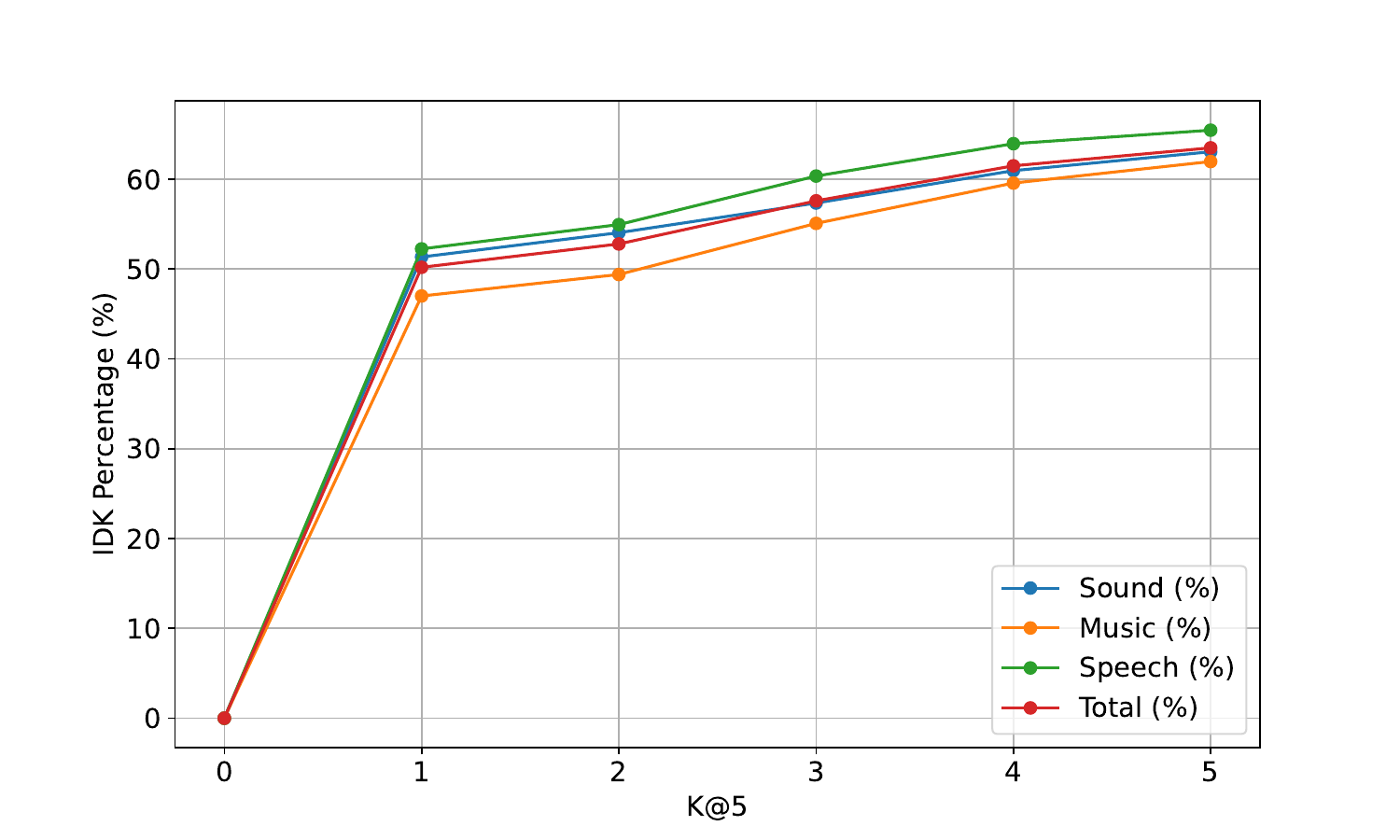}
  \caption{IDK percentage for constructing IDK dataset with different $K$@$5$ threshold.}
  \label{fig:idk_dataset}
\end{figure}

Figure~\ref{fig:idk_dataset} illustrates the percentage for constructing IDK dataset using different $K$@$N$ thresholds. In our experiments, $N$ is set to 5. 
As the $K$@$5$ threshold increases, the requirement for model’s certainty becomes stricter, resulting in a higher percentage of IDK data. 
However, the change in IDK percentage from $1$@$5$ = $50.2\%$ to $5$@$5$ = $63.5\%$ is relatively small, compared to the text modality~\cite{cheng2024can}. 
This suggests that although the capability of LALMs still requires improvement compared to the text modality, their response stability is relatively high, indicating a strong foundation to develop reliable methods. 

The selection of LoRA weights is crucial for balancing between helpfulness and truthfulness. 
Figure~\ref{fig:lora_alpha} illustrates the impact of various LoRA alpha weights on Rejection Rate (\%), Reliability (\%), and Reliability Gain Index (RGI). The Rejection Rate for the main results is provided in Appendix~\ref{sec:rejection_rate_on_different_modalities}. 
The model undergoes SFT for reliability on the speech modality and is tested on the sound and music modalities. 
As shown in Figure~\ref{fig:lora_alpha}(a), the Rejection Rate increases with the LoRA alpha weight grows, indicating that a smaller LoRA alpha weight prevents the model from learning to reject unknown answers, while a larger LoRA alpha weight leads to over-conservatism. 
In Figure~\ref{fig:lora_alpha}(b), the Reliability metric initially increases with the LoRA alpha weight but eventually decreases, demonstrating a non-monotonic relationship. 
Figure~\ref{fig:lora_alpha}(c) shows that the RGI value decreases as the LoRA alpha weight grows, reaching a point when the training becomes ineffective (when RGI < 0). Interestingly, very small LoRA alpha weights can also achieve high RGI values, suggesting that the awareness of reliability is relatively easy to acquire and transfer across modalities. 

\section{Conclusion \& Future Work}
In this work, we have systematically investigated the reliability of large audio language models (LALMs), introducing both training-free and training-based methods to reject questions the LALM cannot answer. 
We propose the novel Reliability Gain Index (RGI) metric, which quantifies the effectiveness of different reliable methods in improving model reliability. 
We have demonstrated that awareness of reliability is a ``meta ability'' of the model, and this awareness can be transferred across various audio modalities, including speech, sound, and music, even when these modalities differ significantly in structure and content. 
Our findings contribute to the ongoing efforts to build more reliable LALMs and provide a foundation for future work in this direction. 
While our study has investigated the transferability of reliability awareness across different audio modalities, future work will explore the possibility of transferring this capability between even more disparate modalities, such as the speech and video modalities, within the context of an Omni Language Model (OLM). 

\section*{Limitation}
While this work has made significant strides in investigating the reliability of LALMs by focusing on their ability to reject questions with ``I don't know'', it primarily addresses the basic aspect of model reliability. 
Specifically, our study does not explore the potential for the model to provide more detailed justifications for its refusal. 
A promising direction for future research is to enable models to actively ask for additional information in an interactive manner when they are unsure, and to provide more reliable answers based on a deeper understanding of the user's query. 
This would not only enhance the model's reliability awareness but also make it more context-aware and capable of engaging in dynamic interactions with users, ultimately leading to more intelligent and trustworthy responses.

\bibliography{custom}

\newpage
\appendix

\section{Dataset Details}
\label{sec:dataset_details}
MMAU~\cite{sakshi2024mmau} is a novel benchmark designed to evaluate the capabilities of large-scale multimodal audio understanding models. 
MMAU consists of $10,000$ carefully curated audio-question-answer pairs, covering three major audio domains: speech, sound, and music. 
These questions involve both information extraction and reasoning tasks, spanning $27$ distinct skills that challenge models to demonstrate advanced audio perception and domain-specific reasoning abilities. 
The dataset is divided into two parts: the Test-mini set, containing $1,000$ questions, and the main Test set, which includes $9,000$ questions. 
As the Test set is not open-sourced, we used the Test-mini set for our experiments. 
The Test-mini set reflects the same task distribution as the main Test set, and thus serves as a reliable evaluation set for reliable methods. 

\begin{table*}[h]
\centering
\caption{\textit{Accuracy} (Acc\%$\uparrow$), \textit{Truthfulness} (Tru\%$\uparrow$), and \textit{Reliability} (Rel\%$\uparrow$) performance comparison of Qwen-Audio-Chat, SALMONN, and Qwen2-Audio-Instruct on the MMAU benchmark across sound, speech, and music modalities. Both the baseline and the IDK Prompting approaches were evaluated, with GPT answer normalization applied. The best-performing items are highlighted in \textbf{bold}, and the second-best items are \underline{underlined}. }
\label{tab:other_lalm_results}
\resizebox{\linewidth}{!}{
\begin{tabular}{lccccccccccccc}
\toprule \toprule 
\multirow{2.5}{*}{\textbf{Models}} & \multirow{2.5}{*}{\makecell{\textbf{IDK} \\ \textbf{Prompting}}} & \multicolumn{3}{c}{\textbf{Sound}} &  \multicolumn{3}{c}{\textbf{Music}} &  \multicolumn{3}{c}{\textbf{Speech}} & \multicolumn{3}{c}{\textbf{Total}} \\ 
\cmidrule{3-14}
& & Acc\% & Tru\% & Rel\% & Acc\% & Tru\% & Rel\% & Acc\% & Tru\% & Rel\% & Acc\% & Tru\% & Rel\% \\
\midrule \midrule

Qwen-Audio-Chat & \CROSS & 57.66 & 58.86 & 58.84 & 53.29 & 53.59 & 53.59 & 35.44 & 35.74 & 35.73 & 48.80 & 49.40 & 49.40 \\
 
Qwen-Audio-Chat & \CHECK & 53.75 & 55.26 & 55.23 & 53.29 & 53.59 & 53.59 & 37.24 & 37.54 & 37.54 & 48.10 & 48.80 & 48.80 \\

\cdashline{1-14}

SALMONN & \CROSS & 50.46 & 52.25 & 52.22 & 49.70 & 50.00 & 50.00 & 30.63 & 37.54 & 37.06 & 43.60 & 46.60 & 46.51 \\

SALMONN & \CHECK &  28.53 & \textbf{87.69} & 52.69 & 17.96 & \textbf{85.03} & 40.05 & 8.41 & \textbf{92.79} & 21.59 & 18.30 & \textbf{88.50} & 39.22 \\

\cdashline{1-14}

Qwen2-Audio-Instruct & \CROSS & \textbf{60.96} & 60.96 & \underline{60.96} & \textbf{55.09} & 55.09 & \underline{55.09} & \textbf{50.75} & \underline{50.75} & 50.75 & \textbf{55.60} & 55.60 & \underline{55.60} \\

Qwen2-Audio-Instruct & \CHECK & \underline{58.26} & \underline{76.28} & \textbf{73.03} & \underline{54.19} & \underline{66.77} & \textbf{65.19} & \underline{43.84} & \underline{58.26} & \textbf{56.18} & \underline{52.10} & \underline{67.10} & \textbf{64.85} \\

\bottomrule \bottomrule
\end{tabular}
}
\end{table*}

\section{Model Details}
\subsection{Introduction for different LALMs}
\label{sec:introduction_for_different_LALMs}

SALMONN\footnote{\url{https://huggingface.co/tsinghua-ee/SALMONN/blob/main/salmonn_v1.pth}} \cite{tang2023salmonn} is one of the first universal LALMs capable of understanding and reasoning about speech, music, and general sounds. It employs a dual-encoder architecture, with the Whisper-Large-V2 \cite{radford2023robust} as the speech encoder and the BEATs \cite{chen2022beats} as the audio encoder. The outputs from both encoders are concatenated and processed by a window-level Q-Former \cite{li2023blip} to align with the LLM Vicuna-13B \cite{chiang2023vicuna}. The entire model was trained in three stages: the pre-training stage aimed at bridging the gap between audio encoders and the LLM, followed by instruction-tuning and activation-tuning stages to enhance the model’s ability to follow human instructions and activate zero-shot emergent capabilities.

Qwen-Audio-Chat\footnote{\url{https://huggingface.co/Qwen/Qwen-Audio-Chat}} \cite{chu2023qwen} is a powerful LALM specifically designed to achieve universal audio understanding and facilitate flexible interaction based on human instructions.  Based on Whisper-Large-V2 \cite{radford2023robust} and Qwen-7B \cite{bai2023qwen}, the model underwent a two-stage training process. In the first stage, a multi-task learning framework incorporating over 30 audio-related tasks was employed to endow the model with a comprehensive understanding of audio data. In the second stage, instruction-based fine-tuning was applied to enhance the model's ability to align with human intent, resulting in a strong interactive chat model. 

Qwen2-Audio-Instruct\footnote{\url{https://huggingface.co/Qwen/Qwen2-Audio-7B-Instruct}} \cite{Qwen2-Audio} represents the latest advancement in the Qwen-Audio series, capable of processing diverse audio inputs and providing either audio analysis or direct textual responses based on speech instructions. The model employs Whisper-Large-V3 \cite{radford2023robust} as its audio encoder and has undergone both supervised fine-tuning (SFT) and direct performance optimization (DPO) after pre-training, which has significantly enhanced its ability to follow complex instructions.  Demonstrating strong performance across multiple benchmarks~\cite{yang2024air, li2024omnibench, sakshi2024mmau}, Qwen2-Audio-Instruct is one of the most powerful open-source LALMs currently available.

\subsection{Performance of different LALMs}
\label{sec:performance_of_different_LALMs}
We evaluated the performance of these powerful LALMs on the Test-mini set of MMAU. The baseline prompt in Table \ref{table:baseline_template} and the IDK prompt in Table \ref{table:IDK_prompting_template} are used to examine the effectiveness of the training-free method. 
As shown in Table \ref{tab:other_lalm_results}, Qwen-Audio-Chat exhibits weak instruction-following capabilities, and adding the IDK prompt had little to no impact on accuracy, truthfulness, or reliability, potentially because the data used in the second stage was much smaller than the data used in the pre-training stage. 
In contrast, SALMONN demonstrated strong instruction-following abilities but was overly conservative. After adding the IDK prompt, the model’s accuracy across all three audio modalities significantly decreased, while its truthfulness notably increased, indicating a strong inclination to refuse to answer questions. 
We hypothesize that this over-strong instruction-following ability is related to the activation tuning in the third stage of SALMONN’s training. 
Qwen2-Audio-Instruct outperforms other models on most of the evaluation metrics, for which it is chosen as the baseline for our main experiments. 

\section{Prompting Details}
\label{sec:prompting_details}
\subsection{Prompt Template for LALM}
\label{sec:prompt_template_for_LALM}
Given \{\textit{Audio}\} and \{\textit{Question}\}, we use some templates to generate unnormalized answers (which means that they cannot be used directly for evaluation processing). For the Baseline and LoRA Fine-tuning (Table~\ref{table:baseline_template}), IDK Prompting (Table~\ref{table:IDK_prompting_template}), and MCoT Prompting (Table~\ref{table:MCoT_prompting_template}), LALM only needs to be inferred once, while for Task Agent, the model needs to be inferred multiple times. The specific templates are shown in the tables bellow. 
\begin{table}[htbp]
\centering
\begin{tcolorbox}[title={Baseline}, colback=white, coltitle=black, colbacktitle=white!0]
\textbf{Input}: \\
\{\textit{Audio}\} \{\textit{Question}\} Select one option from the provided choices: \\
\{\textit{Content\_of\_A}\}\\
\{\textit{Content\_of\_B}\}\\
\{\textit{Content\_of\_C}\}\\
\{\textit{Content\_of\_D}\}\\
\textbf{Output}: \\
\{\textit{Answer}\}\\
\end{tcolorbox}
\caption{The Prompt Template for the baseline on MMAU. }
\label{table:baseline_template}
\end{table}

\begin{table}[htbp]
\centering
\begin{tcolorbox}[title={IDK Prompting}, colback=white, coltitle=black, colbacktitle=white!0]
\textbf{Input}: \\
\{\textit{Audio}\} \{\textit{Question}\} Select one option from the provided choices: \\
\{\textit{Content\_of\_A}\}\\
\{\textit{Content\_of\_B}\}\\
\{\textit{Content\_of\_C}\}\\
\{\textit{Content\_of\_D}\}\\
Output `IDK' if you don't know the answer. \\
\textbf{Output}: \\
\{\textit{Answer}\}\\
\end{tcolorbox}
\caption{The Prompt Template for IDK Prompting on MMAU. }
\label{table:IDK_prompting_template}
\end{table}

\begin{table}[htbp]
\centering
\begin{tcolorbox}[title={MCoT Prompting}, colback=white, coltitle=black, colbacktitle=white!0]
\textbf{Input}: \\
\{\textit{Few-shot Examples}\} \\
\{\textit{Audio}\} \{\textit{Question}\} Select one option from the provided choices: \\
\{\textit{Content\_of\_A}\}\\
\{\textit{Content\_of\_B}\}\\
\{\textit{Content\_of\_C}\}\\
\{\textit{Content\_of\_D}\}\\
Let's think step by step. \\
You can first analyze the sound, music, or speech and then answer the question. \\
Output `IDK' if you don't know the answer. \\
\textbf{Output}: \\
\{\textit{Answer}\}\\
\end{tcolorbox}
\caption{The Prompt Template for MCoT Prompting on MMAU. }
\label{table:MCoT_prompting_template}
\end{table}

\begin{table}[htbp]
\centering
\begin{tcolorbox}[title={Task Agent}, colback=white, coltitle=black, colbacktitle=white!0]
\textbf{Input}: \\
\{\textit{Audio}\} Identify the type of audio. Select one option from the provided choices: \\
Sound\\
Music\\
Speech\\
\textbf{Output}: \\
\{\textit{Type}\} \\
\rule{\linewidth}{0.4pt}
\textbf{Input}: \\
\{\textit{Audio}\} What is the \{\textit{Type}\} content? \\
\textbf{Output}: \\
\{\textit{Content}\}\\
\rule{\linewidth}{0.4pt}
\textbf{Input}: \\
\{\textit{Audio}\} The \{\textit{Type}\} content is: \{\textit{Content}\} \\
\{\textit{Question}\} Select one option from the provided choices: \\
\{\textit{Content\_of\_A}\}\\
\{\textit{Content\_of\_B}\}\\
\{\textit{Content\_of\_C}\}\\
\{\textit{Content\_of\_D}\}\\
Output `IDK' if you don't know the answer. \\
\textbf{Output}: \\
\{\textit{Answer}\}\\
\end{tcolorbox}
\caption{The Prompt Template for Task Agent on MMAU. }
\label{table:task_agent_template}
\end{table}

\subsection{Prompt Template for Answer Normalization}
\label{sec:prompt_template_for_answer_normalization}
Although LALM is required to output a unique option, the output is likely diverse due to limited instruction-following capability of LALM. Therefore, we use the \textit{gpt-4o-mini} API to further normalize the answer. The corresponding prompt is shown in Table~\ref{table:answer_normalization_prompting}. 
\begin{table}[htbp]
\centering
\begin{tcolorbox}[title={Answer Normalization}, colback=white, coltitle=black, colbacktitle=white!0]
\textbf{Input}: \\
According to the answer, select one option from the provided choices. \\
The answer is: \{\textit{Answer}\} \\
The choices are: \\
\{\textit{Content\_of\_A}\}\\
\{\textit{Content\_of\_B}\}\\
\{\textit{Content\_of\_C}\}\\
\{\textit{Content\_of\_D}\}\\
IDK \\
Don't output any other information. \\
\textbf{Output}: \\
\{\textit{Answer (Normalized)}\} \\
\end{tcolorbox}
\caption{The Prompt Template for answer normalization on With OpenAI API. }
\label{table:answer_normalization_prompting}
\end{table}

\section{Training Details}
\label{sec:training_details}
Table~\ref{tab:hyper-parameters} shows the hyper-parameters of SFT in different audio modalities, including learning rate, LoRA alpha, LoRA rank, and LoRA target modules. 
\begin{table}[ht]
    \centering
    \caption{Hyper-parameters with LoRA Fine-tuning for each modality on the MMAU dataset. }
    \label{tab:hyper-parameters}
    \begin{tabular}{c|ccc}
        \toprule
         & Sound & Music & Speech \\
        \midrule
        learning rate      & \multicolumn{3}{c}{$3\times10^{-5}$} \\
        LoRA alpha  & $32$ & \multicolumn{2}{|c}{$16$} \\
        LoRA rank & \multicolumn{3}{c}{$8$} \\
        target modules & \multicolumn{3}{c}{\{k\_proj, q\_proj, v\_proj\}} \\
        \bottomrule
    \end{tabular}
\end{table}

\begin{table*}[htbp]
\centering
\caption{\textit{Accuracy} (Acc\%$\uparrow$), \textit{Truthfulness} (Tru\%$\uparrow$), and \textit{Reliability} (Rel\%$\uparrow$) performance of Qwen2-Audio-Instruct with LoRA Fine-tuning on different modalities. }
\label{tab:domain_reliability}
\begin{tabular}{lccccccccc}
\toprule \toprule
\multirow{2.5}{*}{\textbf{Training Modality}} & \multicolumn{3}{c}{\textbf{Sound}} &  \multicolumn{3}{c}{\textbf{Music}} &  \multicolumn{3}{c}{\textbf{Speech}} \\ 
\cmidrule{2-10}
& Acc\% & Tru\% & Rel\% & Acc\% & Tru\% & Rel\% & Acc\% & Tru\% & Rel\%  \\
\midrule \midrule
Sound & - & - & - & 46.71 & 73.05 & 66.11 & 48.95 & 63.66 & 61.50 \\
Music & 62.76 & 70.57 & 69.96 & - & - & - & 46.85 & 60.06 & 58.31 \\
Speech & 60.66 & 72.97 & 71.46 & 55.99 & 68.26 & 66.76 & - & - & -\\
\bottomrule \bottomrule
\end{tabular}
\end{table*}

\begin{table*}[htbp]
\centering
\caption{\textit{Relative Conservativeness Increase} ($\Delta_{Con}\%\downarrow$), \textit{Relative Humbleness Increase} ($\Delta_{Hum}\%\uparrow$) and \textit{Reliability Gain Index} (RGI$\uparrow$) performance of Qwen2-Audio-Instruct with LoRA Fine-tuning on different modalities.}
\label{tab:domain_rgi}
\resizebox{\linewidth}{!}{
\begin{tabular}{lccccccccc}
\toprule \toprule
\multirow{2.5}{*}{\textbf{Training Modality}} & \multicolumn{3}{c}{\textbf{Sound}} &  \multicolumn{3}{c}{\textbf{Music}} &  \multicolumn{3}{c}{\textbf{Speech}} \\ 
\cmidrule{2-10}
& $\Delta_{Con}\%$ & $\Delta_{Hum}\%$ & RGI & $\Delta_{Con}\%$ & $\Delta_{Hum}\%$ & RGI & $\Delta_{Con}\%$ & $\Delta_{Hum}\%$ & RGI \\
\midrule \midrule
Sound &  - & - & - & 15.57 & 23.95 & 0.19 & 12.31 & 21.02 & 0.23 \\
Music & 6.31 & 14.41 & 0.36 &  - & - & - & 10.81 & 15.32 & 0.15 \\
Speech & 7.51 & 16.82 & 0.35 & 9.88 & 18.26 & 0.27 &  - & - & -  \\
\bottomrule \bottomrule
\end{tabular}
}
\end{table*}

\section{Metric Details}
\label{sec:condition_reliability}

Here we analyze the condition for an invalid Reliability metric. 
In the case of a vanilla LALM, suppose that the model simply answers questions based on its existing knowledge with an accuracy of $\alpha$, where $0 \leq \alpha \leq 1$. As Equation~\ref{eq:rel},the original reliability $Rel_{org}$ of the model can be computed as:
\begin{equation}
\begin{aligned}
Rel_{org} &= 0\cdot \alpha + 1\cdot \alpha \\
&=\alpha
\end{aligned}
\end{equation}

Now, consider the case where a reliable method is applied. 
Let $\Delta_{Con}$ and $\Delta_{Hum}$ represent the increase in conservativeness and humbleness, respectively, as per Equations~\ref{eq:delta_con} and \ref{eq:delta_hum}. 
The lower bound of a valid reliable method occurs when $\Delta_{Con}$ and $\Delta_{Hum}$ are at the same ratio $\rho$. 
The resulting Accuracy, Rejection Rate, Truthfulness, and Reliability can be computed as follows:

\begin{equation}
\begin{aligned}
Acc_{new} &= (1-\rho)\alpha \\
&=\alpha - \rho\alpha
\end{aligned}
\end{equation}

\begin{equation}
\begin{aligned}
Rej_{new} &= \rho\alpha + \rho(1-\alpha) \\
&=\rho
\end{aligned}
\end{equation}

\begin{equation}
\begin{aligned}
Tru_{new} &= Acc + Rej \\
&=\alpha + \rho - \rho\alpha
\end{aligned}
\end{equation}

\begin{equation}
\begin{aligned}
Rel_{new} &= Rej \cdot Acc + (1-Rej) \cdot Tru \\
&=\rho(\alpha - \rho\alpha) + (1-\rho)(\alpha + \rho - \rho\alpha) \\
&=\alpha - \rho\alpha + \rho - \rho^2
\end{aligned}
\end{equation}

We now analyze when the reliability after applying the method exceeds the original reliability, which leads to the following inequality:

\begin{equation}
\begin{aligned}
& Rel_{new} > Rel_{org} \\
\Rightarrow & \alpha - \rho\alpha + \rho - \rho^2 > \alpha \\
\Rightarrow & \rho <  1 - \alpha
\end{aligned}
\end{equation}
where the Reliability metric does not accurately describe the nature of what it expresses, because ineffective reliable methods increase the Reliability metric if the value of $\rho$ less than $1 - \alpha$ is satisfied. 

For an illustrative example, consider a model initially producing $50\%$ correct and $50\%$ incorrect answers.
By applying Equation~\ref{eq:rel}, we can calculate the Reliability of the original unreliable model as:
\begin{equation}
Rel = 0 \times 50\% + 1 \times 50\% = 50\%.
\end{equation}
After applying a reliable method, if $10\%$ of both the correct and incorrect answers are converted into rejections, the new Accuracy/Rejection/Error rate would be $40\%$/$20\%$/$40\%$, respectively. 
In this case, the reliable method would appear ineffective because, after applying the method, the distribution of correct and incorrect answers turning into rejections is similar to what would occur with random sampling. 
However, the reliability increases to: 
\begin{equation}
Rel = 20\% \times 40\% + 80\% \times 60\% = 56\%,
\end{equation}
indicating an ineffective measurement of the model's reliability.  
Here, the increase in reliability is deceptive, as it results from indiscriminate rejection rather than true improvement. The RGI, introduced in the main text, addresses this issue by comparing the relative increase in humbleness and conservativeness. 

\section{More results}
\subsection{LoRA Fine-tuning Results on Different Modalities}
\label{sec:LoRA_fine-tuning_results_on_different_domains}

Table~\ref{tab:domain_reliability} shows the LoRA fine-tuning performance on the Accuracy, Truthfulness, and Reliability of Qwen2-Audio-Instruct trained on one modality and tested on other modalities, while Table~\ref{tab:domain_rgi} shows the LoRA fine-tuning performance on the Relative Conservativeness Increase, Relative Humbleness Increase, and RGI. 

\subsection{Rejection Rate on Different Modalities}
\label{sec:rejection_rate_on_different_modalities}
Table~\ref{tab:rej_rate} shows the proportion of IDK items in the IDK training dataset and the Rejection Rate tested on different audio modalities. The hyper-parameters come from Table~\ref{tab:hyper-parameters}. 

\begin{table}[htbp]
\centering
\caption{Rejection Rate on different modalities}
\label{tab:rej_rate}
\resizebox{1\linewidth}{!}{%
\begin{tabular}{lcccc}
\toprule
\multirow{2.5}{*}{\makecell{\textbf{Training} \\ \textbf{Modality}}} & \multicolumn{4}{c}{\textbf{Rejection Rate (\%)}} \\
\cmidrule(lr){2-5}
 & \textbf{IDK Dataset} & \textbf{Sound} & \textbf{Music} & \textbf{Speech} \\
\midrule
Sound & 63.06 & 37.24 & 26.35 & 14.71 \\
Music & 61.98 & 7.81 & 34.73 & 13.21 \\
Speech & 65.47 & 12.31 & 12.28 & 56.76 \\
\bottomrule
\end{tabular}%
}
\end{table}

\end{document}